\documentclass[usenatbib,useAMS,usegraphicx]{mn2e}

\title[X-ray -- radio correlation in the Vela pulsar]{Toward explanation of the X-ray -- radio correlation in the Vela pulsar}

\author[S. A. Petrova]{S. A. Petrova
\thanks{E-mail: petrova@ri.kharkov.ua}\\
Institute of Radio Astronomy, NAS of Ukraine, 4, Chervonopraporna
Str., 61002 Kharkov, Ukraine}   
   
\begin{document}

\date{Received\dots}

\pagerange{\pageref{firstpage}--\pageref{lastpage}} \pubyear{2009}

\maketitle

\protect\label{firstpage}

\begin{abstract}
Recent observations of the Vela pulsar have revealed a peculiar connection of its emission in the soft X-ray and radio ranges. We suggest the model of the radio pulse formation in the Vela pulsar, develop the theory of the radio photon reprocessing to high energies and on this basis interpret the observed X-ray -- radio connection. The processes of spontaneous and induced scattering of radio waves off the spiraling particles and their observational consequences are examined. The particles are assumed to acquire relativistic gyration energies due to resonant absorption of the radio emission in the outer magnetosphere of a pulsar. The spectral and angular distributions of the spontaneously scattered power are analyzed and compared with the characteristics of the particle synchrotron emission. The consequences of intensity transfer from the radio beam to the background in the course of induced scattering are studied as well. It is demonstrated that the induced scattering can account for the basic features of the Vela's radio profile and its pulse-to-pulse fluctuations. In particular, it can explain a greater role of the leading component and its earlier arrival in stronger pulses. The studies of the radio photon reprocessing to high energies in application to the Vela pulsar shows that the scattered and synchrotron spectra peak at $\sim 0.8$ keV and $0.2$ keV, respectively, with the corresponding luminosities of $10^{29}$ erg\,s$^{-1}$ and $10^{31}$ erg\,s$^{-1}$. The reprocessing inside the light cylinder results in the profile component which stretches from the radio pulse position to $\sim 30^\circ$ earlier in pulse phase. The synchrotron emission beyond the light cylinder presumably contributes to the component which lags the radio pulse by $\sim 90^\circ$. Within the framework of our model, the observed X-ray -- radio connection is explained in terms of the interplay between the processes of induced and spontaneous scattering of the radio pulse.
\end{abstract}


\section{Introduction}

The radio emission of pulsars and their non-thermal high-energy emission have essentially different properties. They are undoubtedly generated by the distinct mechanisms, which are still a matter of debate \citep[see, e.g.,][for a review]{h08,m04}. At the same time, the recent observations \citep{s03,l07} have demonstrated that the high-energy profiles can be influenced by the radio pulse intensity. In the Crab pulsar, the optical pulses coincident with the giant radio pulses are 3\% brighter than the average pulse \citep{s03}. In the Vela pulsar, the shape of the X-ray profile changes with the radio pulse intensity \citep{l07}: the stronger the radio pulse (the earlier it arrives), the more pronounced is the main X-ray peak; besides that, during the weak radio pulses (which arrive later) the excess X-ray flux shifts to the 'trough', which follows the second X-ray peak and spans the longitudinal interval close to the radio pulse window.

Most of the high-energy emission mechanisms discussed in the literature do not depend on the radio pulse intensity. One of the mechanisms, however, suggests the radio photon reprocessing into the high-energy band and therefore implies a physical connection between the radio and high-energy emissions of pulsars \citep{lp98,g01,p03,hum05,hdf08}. In the outer magnetosphere of a pulsar, the radio emission is subject to resonant absorption by the secondary plasma particles. As a result, the absorbing particles acquire substantial transverse momenta \citep{lp98}. In case of strong enough absorption, the longitudinal momenta can also increase significantly \citep{p02}. Then the spontaneous synchrotron radiation of the absorbing particles falls into the optical and X-ray range. This has been suggested to account for the non-thermal high-energy emission of the young and middle-aged pulsars \citep{p03}. Later on the model was extended to the millisecond pulsars by including the influence of the non-compensated accelerating electric field on the particle momentum evolution \citep{hum05}.

Higher intensities of the optical pulses observed in the Crab pulsar during giant radio pulses \citep{s03} may well be explained in terms of synchrotron re-emission of the absorbing particles. Resonant absorption of giant radio pulses should lead to a more pronounced momentum evolution of the particles and, correspondingly, to the larger synchrotron power emitted. At the same time, the intensity redistribution in the high-energy profile of the Vela pulsar  and its dependence on the radio pulse intensity \citep{l07} still remains puzzling. This problem is addressed in the present paper. Firstly, we explain the radio profile structure of the Vela pulsar and its intensity-dependent variations. Secondly, we consider one more mechanism, which deposits the radio photons into the high-energy band, and interpret the high-energy profile. Finally, the observational manifestations of the X-ray -- radio connection are analyzed.

The radio observations of the Vela pulsar \citep{kd83} have revealed that its profile structure depends on the pulse intensity: in stronger pulses, the leading component (the precursor) is much more pronounced and arrives somewhat earlier, whereas the rest of the pulse (the main pulse) is markedly weaker an retains its position in the pulse window. \citet{kd83} have suggested that the precursor component originates in a separate emission region, at higher altitudes in the magnetosphere. Then its location in the pulse profile can be explained in terms of rotational aberration. However, if the emission regions of the precursor and the main pulse are completely independent, it is difficult to understand the observed intensity redistribution between these components, which is controlled by the total intensity of the pulse.

We suggest that the precursor component results from the induced scattering of the main pulse into the background \citep[see also][]{p08b}. The main pulse emission is scattered off the particles, which participate in the resonant absorption and, consequently, perform relativistic helical motion. The induced scattering transfers the radio intensity from the main pulse to the precursor component. It will be shown that stronger induced scattering entails larger intensities of the resultant radio pulse and also its earlier arrival. Thus, it is the process that can account for the the radio profile structure of the Vela pulsar and its intensity-dependent variations.

The high-energy profile of the Vela pulsar has the following structure \citep{he1,g98,he5,h02,he2,he4,he3}. There are two main peaks separated by about a half of the pulsar period. None of them coincide with the radio pulse, and peak 1 lags it by $\sim 90^\circ$ in pulse phase. In the optical and soft X-ray profiles, peak 1 noticeably shifts with frequency toward the radio pulse position, whereas at higher energies the components keep their positions fixed. \citet{g98} has discovered an additional component of the optical profile, which coincides with the radio pulse and extends to the earlier pulse phases. In the soft X-ray range, the component is also present, and one can discern its two peaks, one of which (peak 3) precedes the radio pulse and another one (peak 4) coincides with it \citep{h02}. The component is most pronounced at a few tenths keV and vanishes above a few keV. In the observations of \citet{l07}, the X-ray profile is integrated over the range of 2--16 keV, and this component looks as a trough, but still interacts with the radio pulse.

In the present paper, we explain the trough at keV energies and its connection to the radio pulse. For this purpose we turn to one more mechanism -- the spontaneous scattering off the spiraling particles, -- which deposits the radio photons into the high-energy range. The radio photon reprocessing to high energies implies a physical connection between the radio and high-energy emissions, which can manifest itself in the simultaneous fluctuations in these ranges. The fluctuations are believed to result from the variations of the physical conditions in the magnetosphere. In our model, the radio pulse participates in both the spontaneous and induced scatterings, and it will be shown that the interplay between these processes in the course of the pulse-to-pulse fluctuations of the plasma parameters can account for the X-ray -- radio connection observed in the Vela pulsar.

The plan of the paper is as follows. Section 2 is devoted to the theory of spontaneous scattering off the spiraling particles. The spectral and angular distributions of the scattered power are examined and compared with those of the synchrotron radiation of the scattering particles. The details of induced scattering of the radio waves below the resonance are given in Sect.~3. In Sect.~4 we apply our formalism to the Vela pulsar. The radio profile formation is considered in Sect.~4.1, the high-energy emission is addressed in Sect.~4.2, and the observational manifestations of the X-ray -- radio connection are investigated in Sect.~4.3. Our results are discussed and summarized in Sect.~5.

\section{Spontaneous scattering off the spiraling particles \protect\label{s2}}

The cross-section for the magnetized scattering by the particle at rest was first obtained in \citet{c71}. In application to the pulsar magnetosphere, the scattering in a strong magnetic field by the particles streaming relativistically along the magnetic lines was examined in \citet{bs76,lp96}. \citet{wr78} have considered the non-magnetized induced scattering in the pulsar wind.

Let us consider the radio wave scattering off the particles performing relativistic helical motion in the magnetic field of a pulsar. Deep in the magnetosphere the particles stream relativistically along the open magnetic lines. In the vicinity of the radio emission region, the magnetic field is so strong that any perpendicular momentum of the particles is almost immediately lost via synchrotron re-emission and the radio wave frequency in the particle rest frame is much less than the electron gyrofrequency, $\omega\eta\gamma\ll\omega_G\equiv eB/mc$ (here $\eta\equiv 1-\beta\cos\theta$, $\beta$ is the particle velocity in units of $c$, $\theta$ is the wavevector tilt to the magnetic field, $\gamma$ is the particle Lorentz-factor, $\gamma\equiv(1-\beta^2)^{-1/2}$).

As the magnetic field strength rapidly decreases with distance from the neutron star, $B\propto r^{-3}$, in the outer magnetosphere the radio waves pass through the cyclotron resonance, $\omega\eta\gamma=\omega_G$, where they are subject to resonant absorption. As a result of this process, the incident radio emission is partially absorbed and the particles acquire transverse momenta. In the resonance region, the magnetic field is weak enough, and the spontaneous synchrotron re-emission does not prevent the momentum growth. As is shown in \citet{p02,p03}, the particle gyration becomes relativistic at the very bottom of the resonance region, and further on the transverse and total momenta of the particles continue growing.

Pulsar radio emission is essentially broadband and, correspondingly, the resonance region is sufficiently extended. Over most part of this region there is a significant amount of the photons with frequencies well below the resonance, $\omega\eta\ll\Omega\equiv eB/\gamma mc$. We are interested in the scattering of the under-resonance radio emission off the relativistic spiraling particles. In our case the incident radiation presents the transverse electromagnetic waves polarized either in the plane of the ambient magnetic field (A-polarization) or perpendicularly to this plane (B-polarization). 

The scattering by the electron on a circular orbit has been examined in \citet{p08a}. It has been shown that the under-resonance waves are predominantly scattered to high harmonics of the particle gyrofrequency and may contribute to the observed high-energy emission. (It has also been shown that the effect of the scattering on the particle momenta is negligible.) In the present paper, we extend this formalism to the case of relativistic helical motion of the scattering particles, examine the spectral and angular distributions of the scattered radiation and compare them with those of the synchrotron radiation of the same particles.

Given that the incident radio frequency is well below the resonance, the components of the differential scattering cross-sections at the non-zero harmonics of the gyrofrequency, $s\neq 0$, for different polarization channels are given by Eq.~(16) in \citet{p08a}. The relativistic transformation of the cross-section reads
\begin{equation}
 \frac{{\mathrm d}\sigma}{\mathrm{d}O^\prime}=\left (\frac{{\mathrm d}\sigma}{\mathrm{d}O^\prime}\right )_c\frac{\eta^2}{\eta^{\prime^3}\gamma_\Vert^2},
 \label{eq0}
 \end{equation}
where $\gamma_\Vert$ is the Lorentz-factor of the longitudinal motion, the primes denote the characteristics of the scattered radiation and the subscript 'c' refers to the guiding-centre frame. With Eq.~(\ref{eq0}), the cross-section components for the case of relativistic helical motion of the scattering particles are written as
\[
\frac{{\mathrm d}\sigma_s^{AA}}{{\mathrm d} O^\prime}=\frac{r_e^2}{\gamma^2\eta^{\prime^5}}\frac{\Omega^4}{\omega^4\eta^4} s^4J_s^2\left (\frac{ s\beta_\perp\sin\theta^\prime}{\eta^\prime}\right )\]
\[\times\frac{\sin^2\theta}{\sin^2\theta^\prime}\left (\cos\theta^\prime-\beta_\Vert\right )^4,
\]
	\[
\frac{{\mathrm d}\sigma_s^{AB}}{{\mathrm d} O^\prime}=\frac{r_e^2}{\gamma^2\eta^{\prime^5}}\frac{\Omega^4}{\omega^4\eta^4} s^4J_s^{\prime 2}\left (\frac{ s\beta_\perp\sin\theta^\prime}{\eta^\prime}\right )\]
\[\times\beta_\perp^2\sin^2\theta\left (\cos\theta^\prime-\beta_\Vert\right )^2,
\]
\[
\frac{{\mathrm d}\sigma_s^{BA}}{{\mathrm d} O^\prime}=\frac{r_e^2}{\gamma^2\eta^{\prime^5}}\frac{\Omega^2}{\omega^2\eta^2} s^4J_s^{2}\left (\frac{ s\beta_\perp\sin\theta^\prime}{\eta^\prime}\right )\frac{\left (\cos\theta^\prime-\beta_\Vert\right )^2}{\eta^2\sin^2\theta^\prime}
\]
\[
\times\left\{\sin\theta\left [\eta\eta^\prime-\beta_\perp^2(1-\cos\theta\cos\theta^\prime)/2\right ]+\eta^2\sin\theta^\prime\cos\Delta\phi\right\}^2,
\]
\[
\frac{{\mathrm d}\sigma_s^{BB}}{{\mathrm d} O^\prime}=\frac{r_e^2}{\gamma^2\eta^{\prime^5}}\frac{\Omega^2}{\omega^2\eta^2} s^4J_s^{\prime 2}\left (\frac{ s\beta_\perp\sin\theta^\prime}{\eta^\prime}\right )\frac{\beta_\perp^2}{\eta^2}
\]
\begin{equation}
\times\left\{\sin\theta\left [\eta\eta^\prime-\beta_\perp^2(1-\cos\theta\cos\theta^\prime)/2\right ]+\eta^2\sin\theta^\prime\cos\Delta\phi\right\}^2,
\label{eq14}
\end{equation}
where $r_e$ is the classical electron radius, $\beta_\Vert$ and $\beta_\perp$ are the components of the particle velocity parallel and perpendicular to the ambient magnetic field, respectively, in units of $c$, $\Delta\phi=\phi-\phi^\prime$ is the difference of the azimuthal wavevector components of the incident and scattered radiation and the superscripts of the cross-sections denote the initial and final polarization states of the waves. Making use of Eq.~(\ref{eq0}) in Eq.~(17) of \citet{p08a} yields the zeroth-harmonic cross-sections
\[
\frac{{\mathrm d}\sigma_0^{AA}}{{\mathrm d} O^\prime}=\frac{r_e^2}{\gamma^2\eta^{\prime^5}}\frac{\sin^2\theta\sin^2\theta^\prime}{\gamma_\Vert^4},
\]
\[
\frac{{\mathrm d}\sigma_0^{AB}}{{\mathrm d} O^\prime}=\frac{r_e^2}{\gamma^2\eta^{\prime^3}}\frac{\omega^2\eta^2}{\Omega^2}
\]
\[
\times\left [(\cos\theta-\beta_\Vert)\cos\Delta\phi-\frac{\beta_\perp^2\sin\theta\sin\theta^\prime(\cos\theta^\prime-\beta_\Vert)}{2\eta^{\prime^2}}\right ]^2,
\]
\[
\frac{{\mathrm d}\sigma_0^{BA}}{{\mathrm d} O^\prime}=\frac{r_e^2}{\gamma^2\eta^{\prime^5}}\frac{\omega^2\eta^2}{\Omega^2}
\]
\[
\times\left [(\cos\theta^\prime-\beta_\Vert)\cos\Delta\phi-\frac{\beta_\perp^2\sin\theta\sin\theta^\prime(\cos\theta-\beta_\Vert)}{2\eta^2}\right ]^2,
\]
\begin{equation}
\frac{{\mathrm d}\sigma_0^{BB}}{{\mathrm d} O^\prime}=\frac{r_e^2}{\gamma^2\eta^{\prime^3}}\frac{\omega^2\eta^2}{\Omega^2}\sin^2\Delta\phi.
\label{eq15}
\end{equation}
The total scattering cross-section has the form
\begin{equation}
\frac{{\mathrm d}\sigma^{ij}}{{\mathrm d}O^\prime}=\frac{{\mathrm d}\sigma_0^{ij}}{{\mathrm d}O^\prime}+2\sum_{ s=1}^\infty\frac{{\mathrm d}\sigma_s^{ij}}{{\mathrm d}O^\prime}.
\label{eq16}
\end{equation}
Comparing Eqs. (\ref{eq14}) and (\ref{eq15}), one can see that the zeroth-harmonic term makes negligible contribution to the total cross-section.

The power scattered by an electron is written as
\begin{equation}
P^{ij}=2I\sum_{ s=1}^\infty\int\limits_0^{2\pi}\int\limits_0^\pi\frac{{\mathrm d}\sigma_s^{ij}}{{\mathrm d}O^\prime}\sin\theta^\prime{\mathrm d}\theta^\prime{\mathrm d}\phi^\prime,
\label{eq17}
\end{equation}
where $I=I(\omega,\theta,\phi)$ is the incident intensity. It is interesting to compare the scattered power with the synchrotron power of the electron,
\[P_{\mathrm{syn}}=\frac{e^2\Omega^2}{c}\sum_{ s=1}^\infty s^2\int\limits_0^\pi\frac{\sin\theta^\prime{\mathrm d}\theta^\prime}{\eta^{\prime^3}}\]
\[\times\left [\left (\frac{\cos\theta^\prime-\beta_\Vert}{\sin\theta^\prime}\right )^2J_s^2\left (\frac{ s\beta_\perp\sin\theta^\prime}{\eta^\prime}\right )+\beta_\perp^2J_s^{\prime 2}\left (\frac{ s\beta_\perp\sin\theta^\prime}{\eta^\prime}\right )\right ],\]
where the first and the second terms correspond to the A- and B-polarizations, respectively. One can observe that $(\partial P_s^{iA}/\partial O^\prime)/(\partial P_s^{iB}/\partial O^\prime)=[(\cos\theta^\prime-\beta_\Vert)^2/\sin^2\theta^\prime]J_s^2/\beta_\perp^2J_s^{\prime^2}$, $(i=A,B)$, similarly to the synchrotron case. However, the angular and spectral distributions of the scattered power are somewhat different.

To examine these distributions in more detail we allow for the relativistic character of the electron gyration and make use of the asymptotic representations of the Bessel function and its derivative with respect to the argument $\xi$ at $\xi\to s-0$:
\[
J_s(\xi)=\frac{\varepsilon^{1/2}}{\pi\sqrt{3}}K_{1/3}\left (\frac{ s}{3}\varepsilon^{3/2}\right ),
\]
\[
J_s^\prime(\xi)=\frac{\varepsilon}{\pi\sqrt{3}}K_{2/3}\left (\frac{ s}{3}\varepsilon^{3/2}\right ),
\]
where $\varepsilon\equiv 1-\xi^2/ s^2$ and $K_\mu(x)$ is the modified Bessel function. It is convenient to introduce the variables
\begin{equation}
\psi=\frac{\beta_0\gamma(\cos\theta^\prime-\beta_\Vert)}{\sqrt{1-\beta_\Vert^2}}\quad \mathrm{and}\quad y=\frac{2 s}{3}\gamma_0^{-3},
\label{eq18}
\end{equation}
where $\beta_0$ is the normalized velocity of the electron in the guiding centre frame and $\gamma_0\equiv(1-\beta_0)^{-1/2}$. The invariance of the transverse momentum implies that $\beta_0\gamma_0=\beta_\perp\gamma$ and, correspondingly, $\gamma_0=\gamma/\gamma_\Vert$. In the case under consideration $\gamma_0\gg 1$ and $\beta_0\approx 1$. As the functions $K_{1/3}(\xi)$ and $K_{2/3}(\xi)$ are significant only for the arguments $\xi\la 1$, one can extend the limits of integration over $\psi$ to $(-\infty,\infty)$. Furthermore, as these functions peak at high harmonics, $ s\sim\gamma_0^3$, one can replace the summation over $ s$ by integration over $y$. Then the scattered power is written as
\[
P^A=4\pi I\frac{r_e^2}{\gamma^2}\frac{\Omega^4}{\omega^4\eta^4}\gamma_0^6\gamma^2\sin^2\theta\frac{81}{32\pi^2}
\]
\[
\times\int\limits_0^\infty y^4\mathrm{d}y\int\limits_{-\infty}^\infty\mathrm{d}\psi\left [\psi^4(1+\psi^2)K_{1/3}^2(\xi)+\psi^2(1+\psi^2)^2K_{2/3}^2(\xi)\right ],
\]
\[
P^B=4\pi I\frac{r_e^2}{\gamma^2}\frac{\Omega^2}{\omega^2}\frac{\gamma_0^6\gamma^4}{2}\left (1+\frac{\sin^2\theta}{2\gamma_\Vert^2\eta^2}\right )\frac{81}{32\pi^2}
\]
\begin{equation}
\times\int\limits_0^\infty y^4\mathrm{d}y\int\limits_{-\infty}^\infty\mathrm{d}\psi\left [\psi^2(1+\psi^2)K_{1/3}^2(\xi)+(1+\psi^2)^2K_{2/3}^2(\xi)\right ],
\label{eq19}
\end{equation}
where the superscripts of $P$  denote the polarization of the incident waves, $\xi\equiv(y/2)(1+\psi^2)^{3/2}$, and the synchrotron power reads
\[P_{\mathrm{syn}}=\frac{e^2\Omega^2\gamma^2\gamma_0^2}{c}\frac{9}{8\pi^2}\]
\begin{equation}
\times\int\limits_0^\infty y^2\mathrm{d}y\int\limits_{-\infty}^\infty\mathrm{d}\psi\left [\psi^2(1+\psi^2)K_{1/3}^2(\xi)+(1+\psi^2)^2K_{2/3}^2(\xi)\right ].
\label{eq20}
\end{equation}

\begin{figure*}
\centering
\includegraphics[width=90mm]{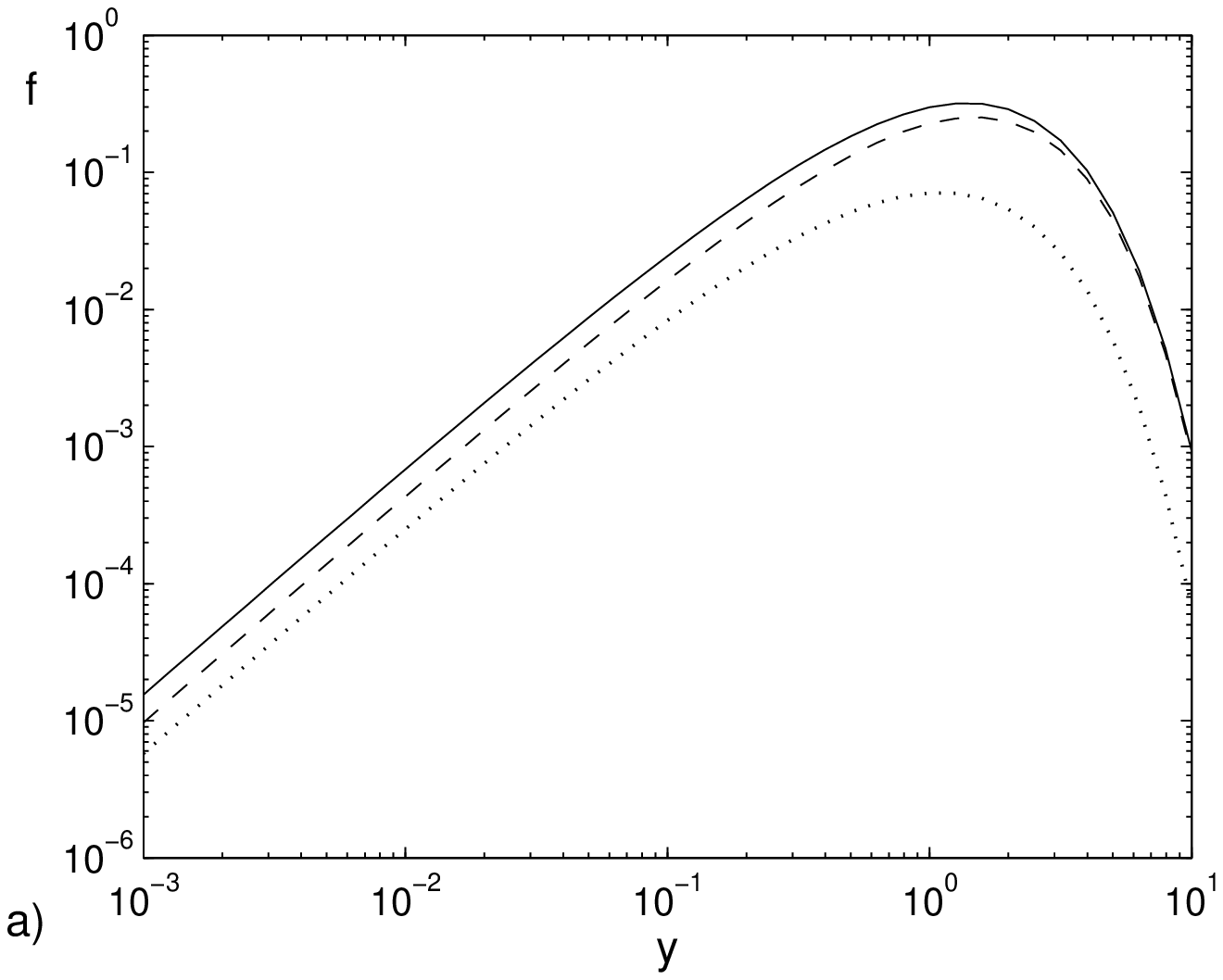}
\includegraphics[width=90mm]{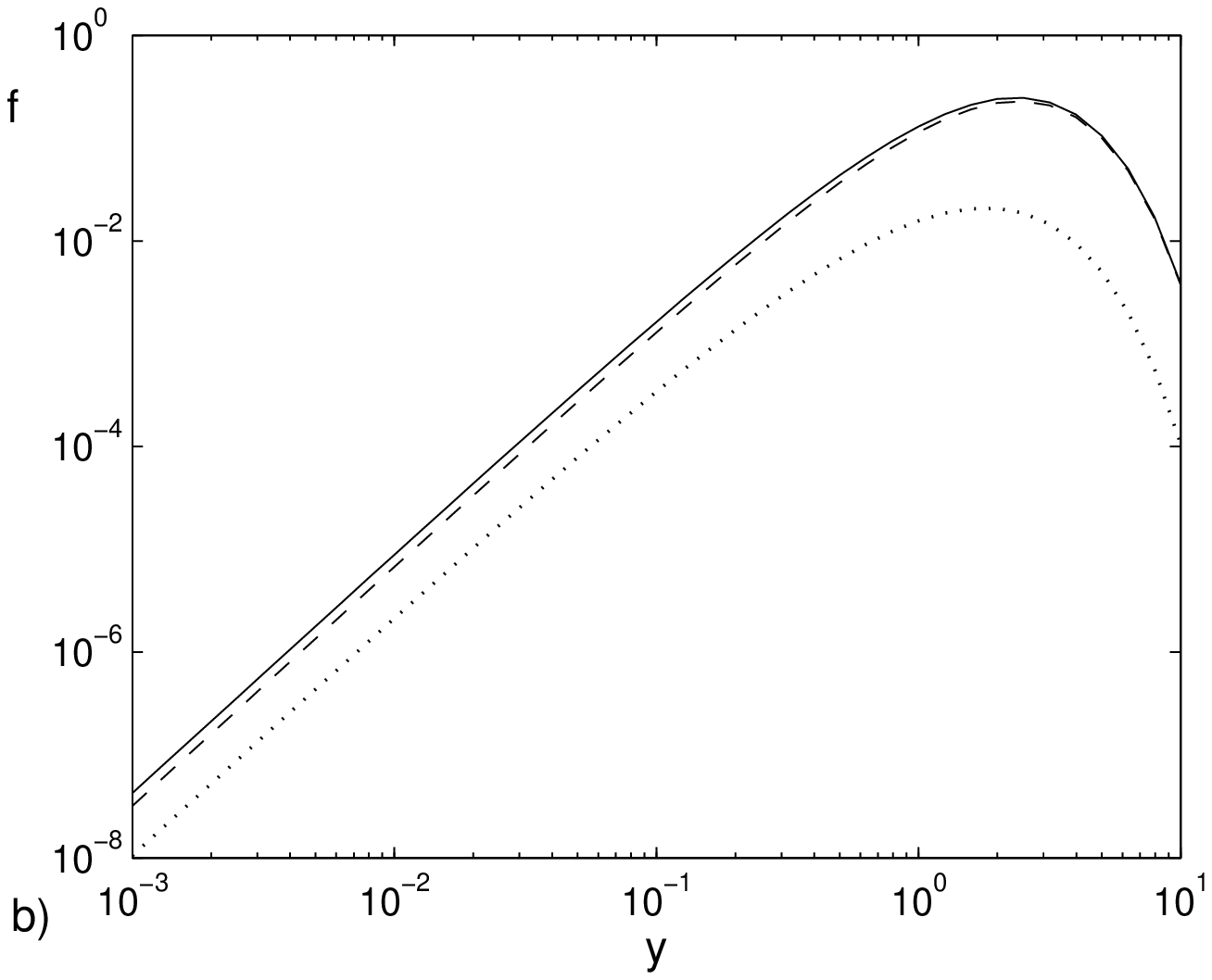}
\includegraphics[width=90mm]{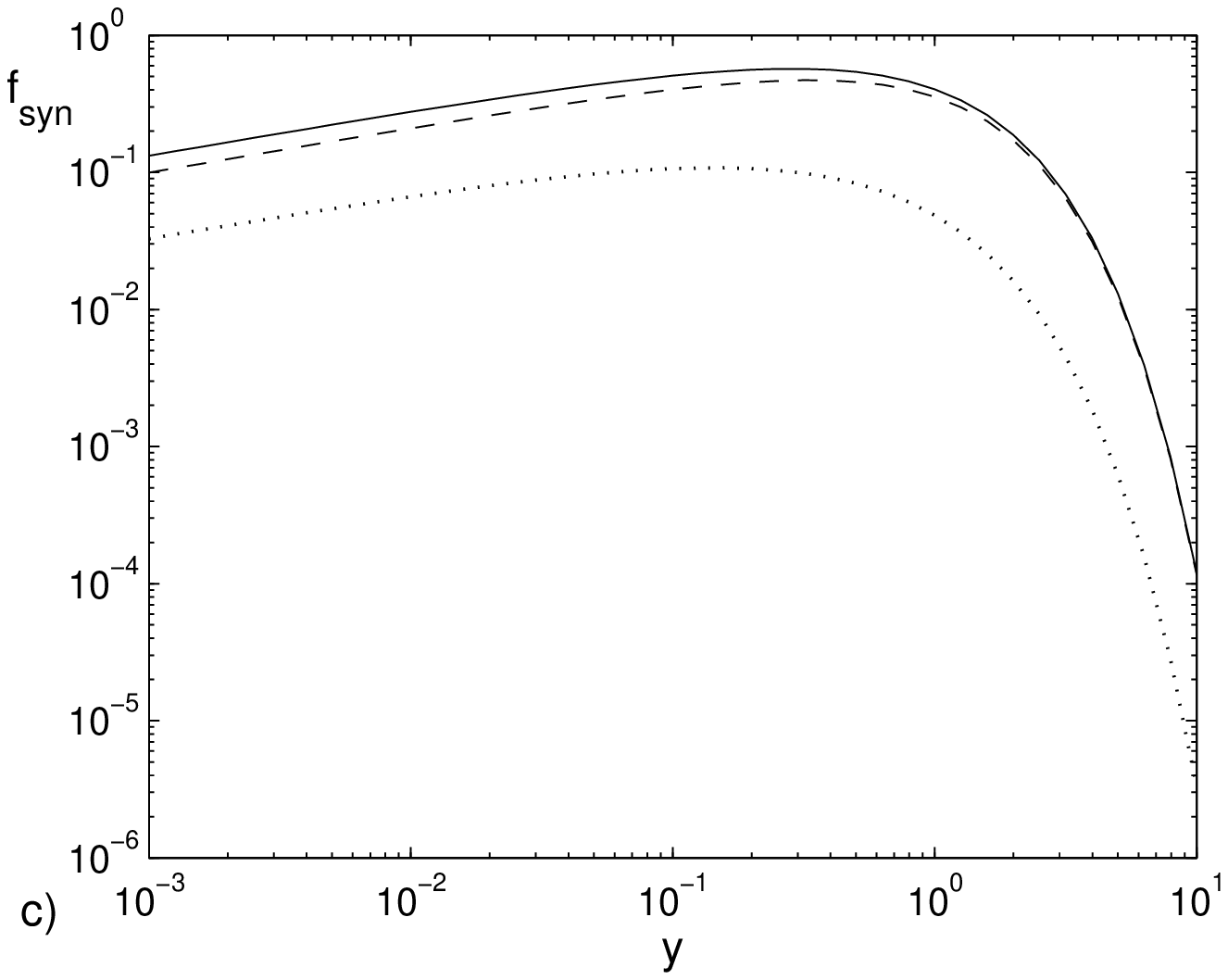}
\caption{Spectral distribution of the power produced by a spiraling particle: {\bf a} -- scattering of the incident A-polarization, {\bf b} -- scattering of the incident B-polarization, {\bf c} -- synchrotron emission; the dotted and dashed lines correspond to the A- and B-polarizations, respectively, and the solid lines show their sum.}
\label{f1}
\end{figure*}

Integrating Eqs.~(\ref{eq19})-(\ref{eq20}) over the angular coordinate $\psi$ with the help of the well-known integrals of the synchrotron theory
\[
\int\limits_{-\infty}^{\infty}\psi^2(1+\psi^2)K_{1/3}^2(\xi)\mathrm{d}\psi=\frac{\pi}{\sqrt{3}y}\left [\int\limits_y^\infty K_{5/3}(x)\mathrm{d}x-K_{2/3}(y)\right ],
\]
\[
\int\limits_{-\infty}^{\infty}(1+\psi^2)^2K_{2/3}^2(\xi)\mathrm{d}\psi=\frac{\pi}{\sqrt{3}y}\left [\int\limits_y^\infty K_{5/3}(x)\mathrm{d}x+K_{2/3}(y)\right ]
\]
and the analogous integrals (\ref{a14}) and (\ref{a18}) calculated in Appendix A, one can obtain the normalized spectral distributions
\[
f^{AA}(y)=\frac{243\sqrt{3}}{256\pi}y^3\left [K_{4/3}(y)-\int\limits_y^\infty K_{5/3}(x)\mathrm{d}x\right ],
\]
\[
f^{AB}(y)=\frac{81\sqrt{3}}{256\pi}y^3\left [\frac{8}{3y}K_{1/3}(y)+K_{4/3}(y)-\int\limits_y^\infty K_{5/3}(x)\mathrm{d}x\right ],
\]
\[
f^{AB}(y)=\frac{81\sqrt{3}}{448\pi}y^3\left [\int\limits_y^\infty K_{5/3}(x)\mathrm{d}x-K_{2/3}(y)\right ],
\]
\[
f^{BA}(y)=\frac{81\sqrt{3}}{448\pi}y^3\left [\int\limits_y^\infty K_{5/3}(x)\mathrm{d}x+K_{2/3}(y)\right ],
\]
\[
f^A_{\mathrm{syn}}(y)=\frac{9\sqrt{3}}{16\pi}y\left [\int\limits_y^\infty K_{5/3}(x)\mathrm{d}x-K_{2/3}(y)\right ],
\]
\begin{equation}
f^B_{\mathrm{syn}}(y)=\frac{9\sqrt{3}}{16\pi}y\left [\int\limits_y^\infty K_{5/3}(x)\mathrm{d}x+K_{2/3}(y)\right ],
\label{eq21}
\end{equation}
where it is taken that $\int_0^\infty(f^{AA}+f^{AB})\mathrm{d}y=\int_0^\infty(f^{BA}+f^{BB})\mathrm{d}y=\int_0^\infty(f^A_{\mathrm{syn}}+f^B_{\mathrm{syn}})\mathrm{d}y=1$. Proceeding from the approximation of the modified Bessel function at small arguments,
\[K_\mu(y)\approx 2^{\mu-1}\Gamma(\mu)y^{-\mu},\quad y\ll 1,\]
where $\Gamma(\mu)$ is the gamma-function, one can find the asymptotic behaviour of the spectra at $y\to 0$:
\[
f^{AA}=\frac{243\sqrt{3}}{256\pi}2^{-5/3}\Gamma(1/3)y^{5/3},\quad f^{AB}=\frac{5}{3}f^{AA},
\]
\[
f^{BA}=\frac{81\sqrt{3}}{448\pi}2^{-4/3}\Gamma(2/3)y^{7/3},\quad f^{BB}=3f^{BA},
\]
\begin{equation}
f^A_{\mathrm{syn}}=\frac{9\sqrt{3}}{8\pi}2^{-4/3}\Gamma(2/3)y^{1/3},\quad f^B_{\mathrm{syn}}=3f^A_{\mathrm{syn}},
\label{eq22}
\end{equation}
Note that the scattered radiation has much steeper spectra. Provided that $y\gg 1$, $K_\mu(y)\approx\sqrt{\pi/2y}\exp(-y)$ and the spectra drop exponentially. The overall spectral distributions (\ref{eq21}) are plotted in Fig.~\ref{f1}. One can see that the power of the scattered radiation peaks at markedly larger frequencies than the synchrotron power in both cases of the incident A- and B-polarizations ($y_{\mathrm{peak}}\approx 1.25$ and $2.5$, respectively, whereas the synchrotron peak lies at $y\approx 0.3$).

Making use of the formula
\[
\int\limits_0^\infty K_\mu(y)y^\rho\mathrm{d}y=2^{\rho-1}\Gamma\left (\frac{\rho+1+\mu}{2}\right )\Gamma\left (\frac{\rho+1-\mu}{2}\right ),\]
\[ \rho+1>\mu,
\]
one can integrate the spectral distributions to obtain the total power in each polarization:
\[
P^{AA}=\frac{\pi r_e^2}{2}I\gamma_0^6\frac{\Omega^4}{\omega^4\eta^4}\sin^2\theta,\quad P^{AB}=\frac{13}{3}P^{AA},
\]
\[
P^{BA}=\frac{2\pi r_e^2}{3}I\gamma_0^6\gamma^2\frac{\Omega^2}{\omega^2}\left (1+\frac{\sin^2\theta}{2\gamma_\Vert^2\eta^2}\right ),\quad P^{BB}=13P^{BA},
\]
\begin{equation}
P^A_{\mathrm{syn}}=\frac{e^2\Omega^2\gamma^2\gamma_0^2}{12c},\quad P^B_{\mathrm{syn}}=7P^A_{\mathrm{syn}}.
\label{eq23}
\end{equation}
It should be noted that the polarization states of the scattered radiation in the cases of incident A- and B-polarizations are distinct and both differ from the synchrotron case. Besides that, $R\equiv(P^{AA}+P^{AB})/(P^{BA}+P^{BB})\sim (\Omega^2/\omega^2\eta^2)(\sin^2\theta/\eta^2\gamma^2)$. Although the quantity $\sin^2\theta/\eta^2\gamma_\Vert^2\gamma_0^2$ may be much less than unity, in our consideration $R\gg 1$, since only the leading terms in $\Omega^2/\omega^2\eta^2$ are retained. Thus, the waves of the A-polarization are scattered much more efficiently. Note also that the linearization technique applied for the derivation of the scattering cross-section (\ref{eq14}) is valid only until the power scattered is less than the synchrotron power of the particle \citep[for more details see][]{p08a}.

\begin{figure*}
\centering
\includegraphics[width=100mm]{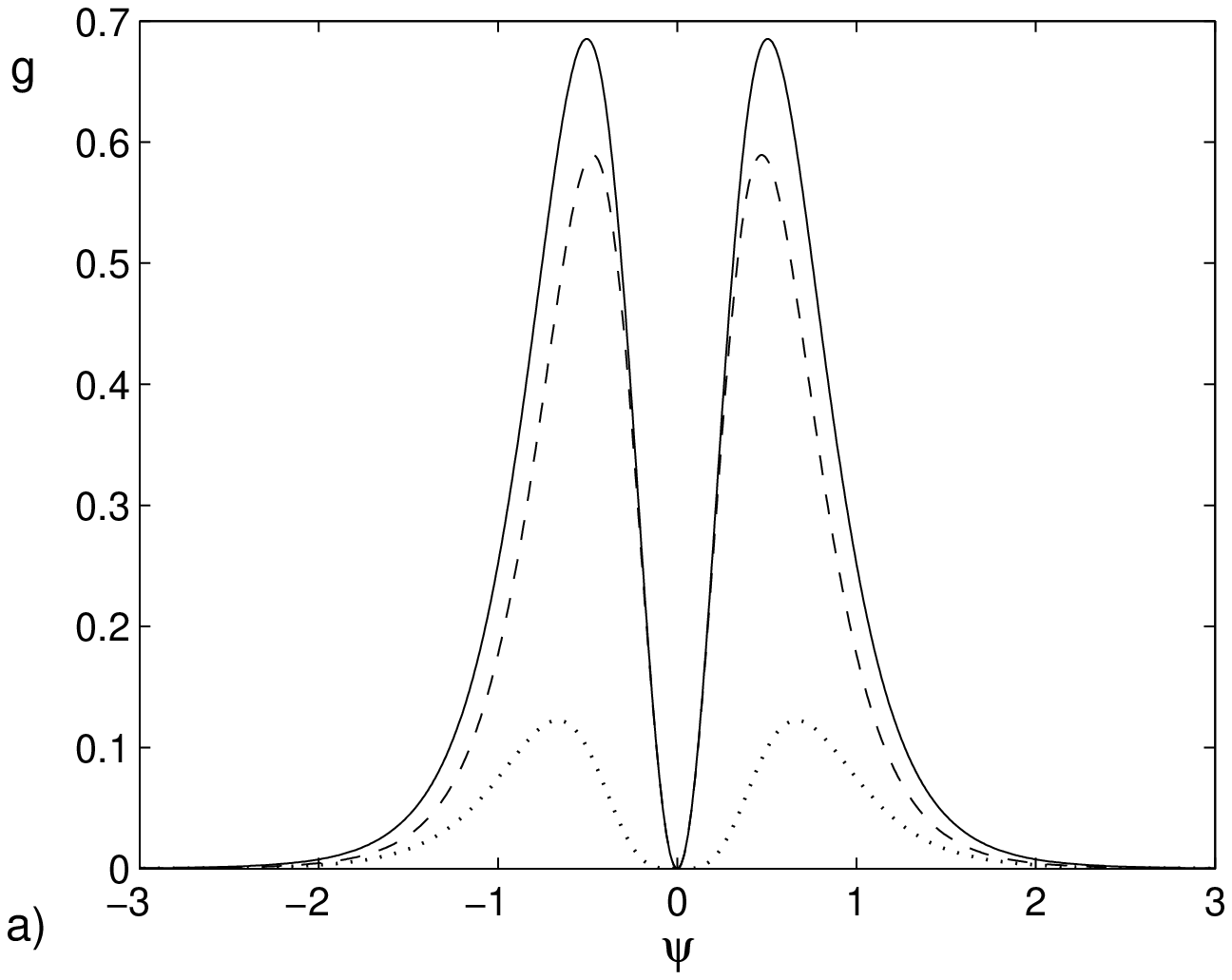}
\includegraphics[width=100mm]{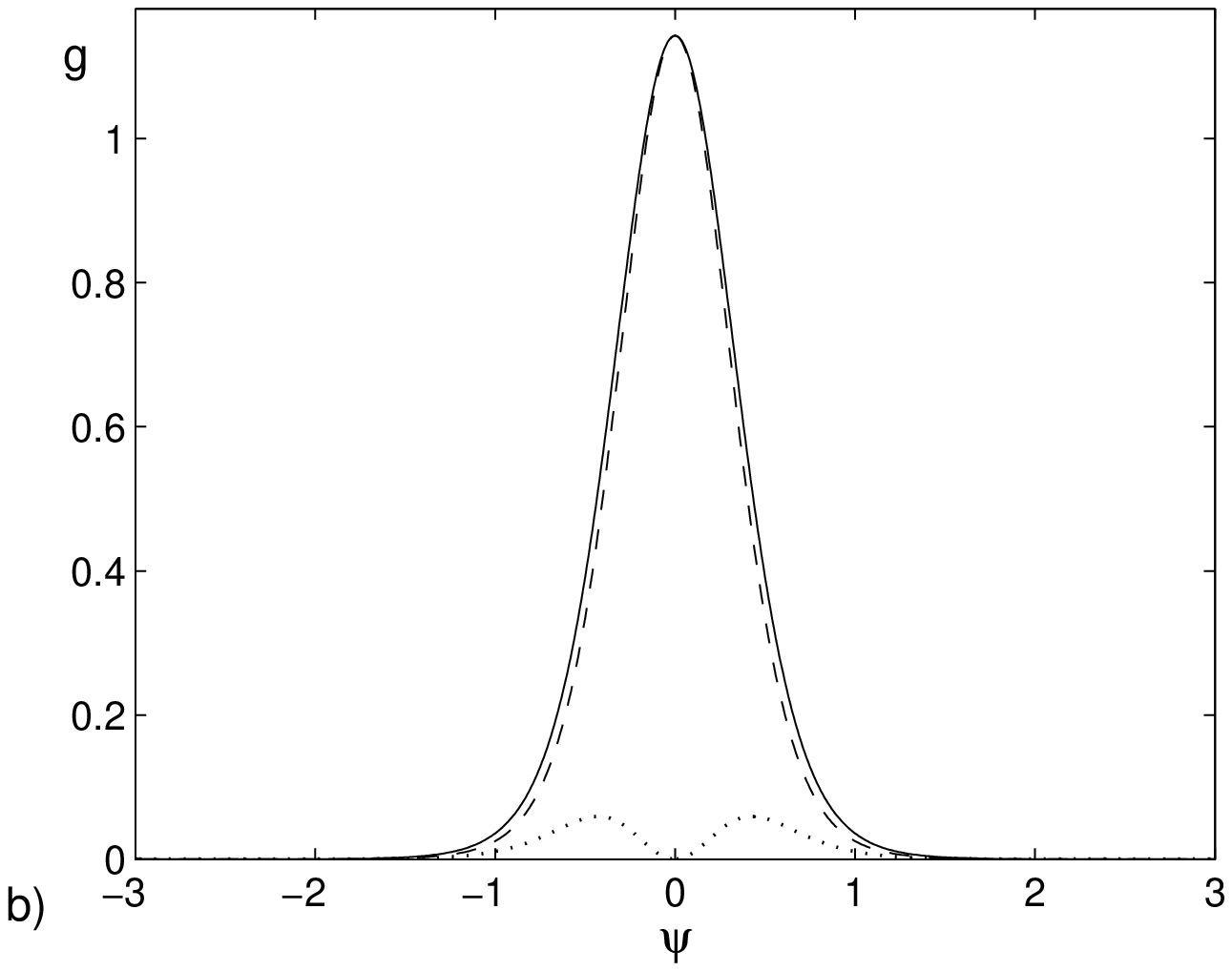}
\includegraphics[width=100mm]{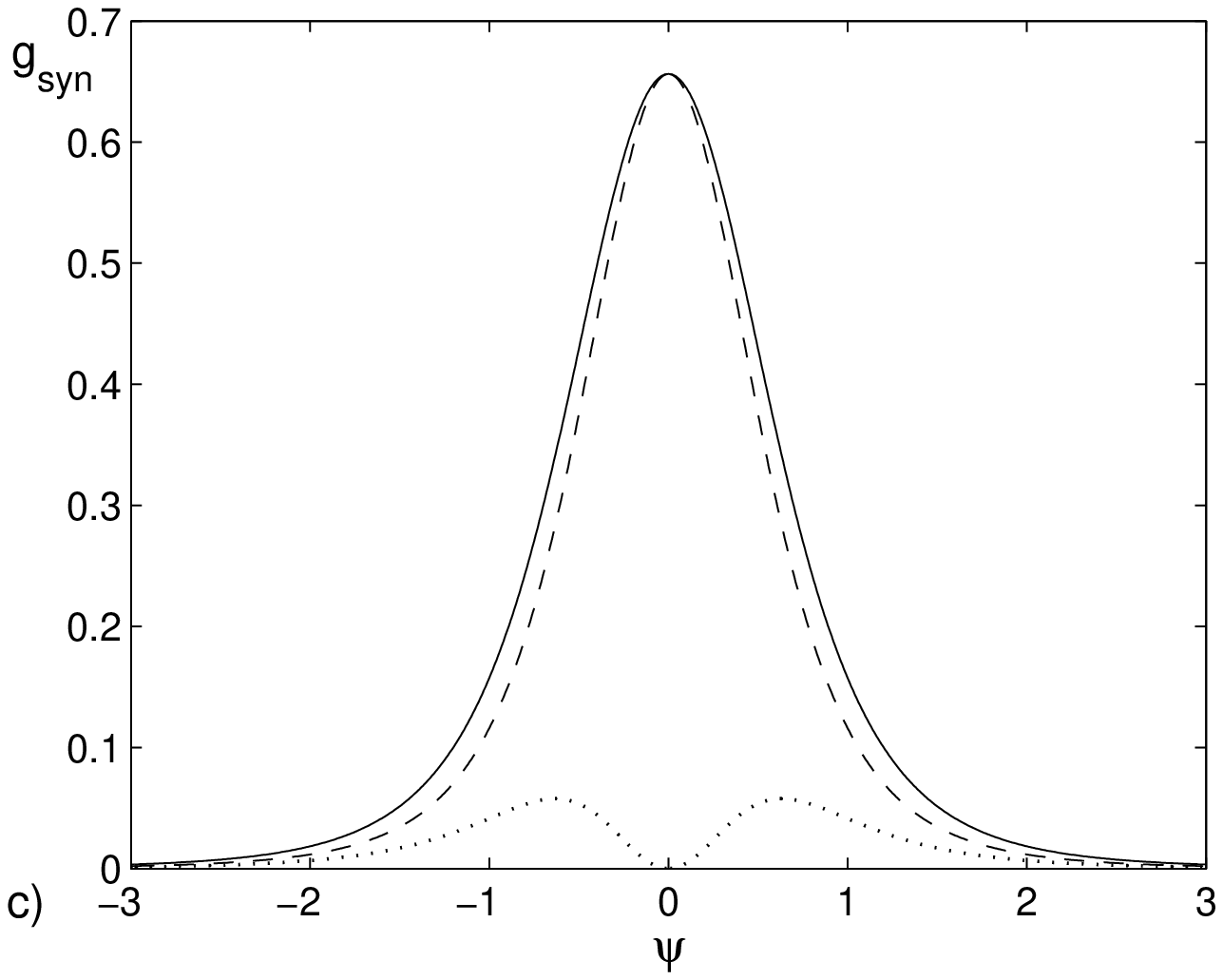}
\caption{Angular distribution of the power produced by a spiraling particle. For details see the caption of Fig.~\ref{f1}.}
\label{f2}
\end{figure*}

The angular distributions can be obtained by integrating Eqs.~(\ref{eq19})-(\ref{eq20}) over $y$ with the help of the integral
\[
\int\limits_0^\infty K_\mu(x)K_\nu(x)x^{\rho-1}\mathrm{d}x=\frac{2^{\rho-3}}{\Gamma(\rho)}\Gamma\left (\frac{\rho+ \nu+\mu}{2}\right )\Gamma\left (\frac{\rho+\nu-\mu}{2}\right )\]
\[\times\Gamma\left (\frac{\rho-\nu+\mu}{2}\right )\Gamma\left (\frac{\rho-\nu-\mu}{2}\right ),\quad \rho>\nu+\mu>0,
\]
and they take the form

\[
g^{AA}=\frac{315}{512}\frac{11\psi^4}{(1+\psi^2)^{13/2}},\quad g^{AB}=\frac{315}{512}\frac{13\psi^2}{(1+\psi^2)^{11/2}},\]
\[
g^{BA}=\frac{45}{512}\frac{11\psi^2}{(1+\psi^2)^{13/2}},\quad g^{BB}=\frac{45}{512}\frac{13}{(1+\psi^2)^{11/2}},\]

\begin{equation}
g^A_{\mathrm{syn}}=\frac{3}{32}\frac{5\psi^2}{(1+\psi^2)^{7/2}},\quad g^B_{\mathrm{syn}}=\frac{3}{32}\frac{7}{(1+\psi^2)^{5/2}},
\label{eq24}
\end{equation}
where $\int_{-\infty}^\infty(g^{AA}+g^{AB})\mathrm{d}\psi=\int_{-\infty}^\infty(g^{BA}+g^{BB})
\mathrm{d}\psi=\int_{-\infty}^\infty(g^A_{\mathrm syn}+g^B_{\mathrm{syn}})\mathrm{d}\psi=1$. These distributions are plotted in Fig.~\ref{f2}. Note the drop of $g^{AA}+g^{AB}$ at $\psi=0$ and the narrowness of the peak of $g^{BA}+g^{BB}$ as compared to the synchrotron case. Integration of the angular distributions using the formula
\[
\int\limits_{-\infty}^{\infty}\frac{\mathrm{d}\psi}{(1+\psi^2)^\rho}=\frac{\Gamma(1/2)\Gamma(\rho-1/2)}{\Gamma(\rho)}
\]
leads again to the total powers given by Eq.~(\ref{eq23}).

\section{Induced scattering off the spiraling particles \protect\label{s3}}

The brightness temperatures of pulsar radio emission are so high that the induced scattering is believed to dominate the spontaneous one. Below we concentrate on the induced scattering of the under-resonance radio waves, $\omega\eta\ll\Omega$. As the magnetic field strength decreases with distance from the neutron star, the wave of a given frequency successfully passes through the resonances of increasingly higher orders. The under-resonance scattering takes place at the lowest altitudes, and it should be most efficient because of larger incident intensity and plasma number density in this region. In the course of induced scattering between the photon states corresponding to different harmonics of the gyrofrequency the photons are transferred from the higher harmonics to the lower ones. Thus, the incident under-resonance waves suffer only the zeroth-harmonic induced scattering, $\omega^\prime\eta^\prime=\omega\eta$.

As is shown in \citet{p08c}, the induced scattering from several first harmonics to the under-resonance state can also be noticeable. At the same time, the induced scattering from the high harmonics, $s\gg 1$, is negligible. Firstly, the spectral intensities of the pulsar high-energy emission are much less than the radio intensities. Besides that, in contrast to the spontaneous scattering, the induced scattering chiefly transfers the photons to the neighbouring harmonics, the process becoming less efficient at $s\gg 1$.

Keeping in mind the above considerations, we are interested in the under-resonance induced scattering, $\omega^\prime\eta^\prime=\omega\eta\ll\Omega$, off the particles performing relativistic helical motion. It should be noted that, according to Eq.~(\ref{eq15}), the scattering between the states with the A-polarization strongly dominates that in the other polarization channels, since only the cross-section $\mathrm{d}\sigma^{AA}/\mathrm{d}O^\prime$ does not contain the factor $\omega^2\eta^2/\Omega^2\ll 1$ . Hence, we dwell on the scattering in the channel $\mathrm{A}\to \mathrm{A}$.

Pulsar radio emission is known to be highly directional. At any point of the emission cone it is concentrated into a narrow beam of the opening angle $\la 1/\gamma_\Vert$, which is typically much less than the angular width of the emission cone. Far enough from the emission region, the radio emission propagates quasi-transversely with respect to the ambient magnetic field, $1/\gamma_\Vert\ll\theta\la 1$.

\begin{figure*}
\centering
\includegraphics{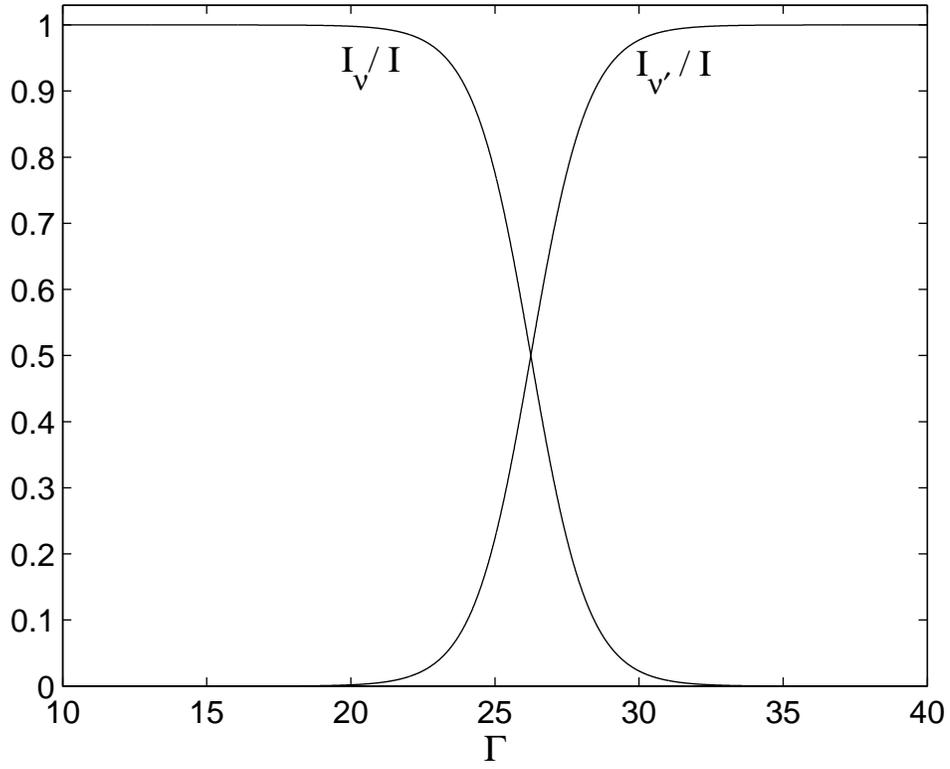}
\caption{Spectral intensities of the radio beam and background as functions of the efficiency of induced scattering; $I_{\nu^\prime}^{(0)}/I_\nu^{(0)}=4\times 10^{-12}$.}
\label{f3}
\end{figure*}

The rate of induced scattering is determined by the particle recoil in the scattering act, and hence, the induced scattering at large angles, out of the radio beam, may be much more efficient than the scattering inside the beam. At the same time, the former process may start only if initially there are some photons outside the beam. Such background photons may be present, e.g., due to the spontaneous scattering from the beam. Although the background photons are very few, they  can still stimulate efficient induced scattering from the beam, and finally a significant part of the beam intensity may be deposited to the background (see Sect.~4.1 below). At the beginning of induced scattering, the background occupation numbers grow exponentially, the exponent being dependent on the photon orientation. Hence, the beam photons are scattered predominantly into the state $\vec{k}^\prime$  corresponding to the maximum scattering probability. Thus, the induced scattering from the radio beam into the background gives rise to a narrow scattered component. As is shown in \citet{p08c}, in the case considered it is directed at the angle $\theta_{\mathrm{max}}^\prime\sim\gamma_0/\gamma$ to the ambient magnetic field. The spectral intensities of the beam and the background evolve as
\begin{equation}
I_\nu=\frac{I/x}{1+1/x},\quad I_{\nu^\prime}=\frac{I}{1+1/x},
\label{eq32}
\end{equation}
where $I\equiv I_\nu+I_{\nu^\prime}=\mathrm{const}$ is the total intensity of the beam and the background, $x\equiv\left [I_{\nu^\prime}^{(0)}/I_\nu^{(0)}\right ]\exp (\Gamma)$, $I_\nu^{(0)}$ and $I_{\nu^\prime}^{(0)}$ are the initial intensities of the beam and the background and 
\begin{equation}
\Gamma=\frac{12IN_er_e^2r}{m\nu^{\prime^2}\gamma\gamma_0^2}.
\label{eq31}
\end{equation}
The quantity $\Gamma$ characterizes the scattering efficiency, whereas $x$ the extent of intensity transfer from the radio beam to the background. As long as $x\ll 1$, the background intensity grows exponentially, $I_{\nu^\prime}\sim I_{\nu^\prime}^{(0)}\exp (\Gamma)$, whereas the beam intensity is almost unaltered, $I_\nu\approx I_\nu^{(0)}$. Given that $x\gg 1$, the background intensity becomes comparable with the initial radio beam intensity, $I_{\nu^\prime}\approx I_\nu^{(0)}$, and enters the stage of saturation, whereas the beam intensity noticeably decreases, $I_\nu\sim I_\nu^{(0)}/x$. The final intensities $I_\nu$ and $I_{\nu^\prime}$ as functions of $x$ are shown in Fig.~\ref{f3}. Note the narrowness of the range of $x$, where both intensities are comparable in magnitude.

\section{Applications to the Vela pulsar \protect\label{s4}}

\subsection{Radio profile formation \label{s4.1}}

Let us examine the radio profile evolution of the Vela pulsar as a result of induced scattering off the spiraling particles. The variations of the conditions in the scattering region should result in the pulse-to-pulse fluctuations of the radio profile, and it is the fluctuations that are believed to be connected with those of the high-energy emission.

Recall that our consideration is restricted to the induced scattering between the photon states with the A-polarization. The radio emission of the Vela pulsar is known to have almost complete linear polarization. It is generally believed that in this pulsar only one polarization mode is present in any radio pulse and at any pulse phase. We assume that this is the A-mode (see Sect.~5 for further discussion).

In order to conclude whether the induced scattering may affect the Vela's radio profile noticeably we start from estimating the level of the background radiation, which results from the spontaneous scattering of the radio pulse: $I_{\nu^\prime}^{(0)}/I_\nu^{(0)}=N_er(\omega/\omega^\prime)\mathrm{d}\sigma/\mathrm{d}O^\prime$. Using Eq.~(\ref{eq15}) and taking into account that $\theta^\prime\approx 1/\gamma_\Vert$ and $\gamma_\Vert=\gamma/\gamma_0$, we obtain
\begin{equation}
\frac{I_{\nu^\prime}^{(0)}}{I_\nu^{(0)}}\approx\frac{2N_er_e^2r}{\gamma_0^2}.
\label{eq34}
\end{equation}
It is convenient to noramlize the number density of the scattering particles by the Goldreich-Julian density,
\begin{equation}
N_e=\frac{\kappa B}{Pce},
\label{eq35}
\end{equation}
where $\kappa$ is the multiplicity factor of the plasma and $P$ is the pulsar period. With the dipolar geometry of the magnetic field, $B\propto r^{-3}$, one can estimate Eq.~(\ref{eq34}) as
\begin{equation}
\frac{I_{\nu^\prime}^{(0)}}{I_\nu^{(0)}}=10^{-11}\frac{0.1\,\mathrm{s}}{P}\frac{B_\star}{10^{12}\,\mathrm{G}}\frac{\kappa}{10^2}\left (\frac{\gamma_0}{10}\right )^{-2}\left (\frac{r}{10^8\,\mathrm{cm}}\right )^{-2},
\label{eq36}
\end{equation}
where $B_\star$ is the magnetic field strength at the neutron star surface and all the quantities are normalized to their characteristic values. For the parameters of the Vela pulsar, $P=0.089$ s and $B_\star=3.4\times 10^{12}$ G, we have $I_{\nu^\prime}^{(0)}/I_\nu^{(0)}\approx 4\times 10^{-12}$. Hence, the induced scattering becomes significant (i.e. $x\sim 1$) for the scattering efficiencies $\Gamma\approx 30$.

To estimate $\Gamma$ we take that $I\equiv I_\nu^{(0)}+I_{\nu^\prime}^{(0)}\approx I_\nu^{(0)}$ and present the radio beam intensity as
\begin{equation}
I_\nu^{(0)}=I_{\nu_0}\left (\frac{\nu}{\nu_0}\right )^{-\alpha},
\label{eq37}
\end{equation}
where $\alpha$ is the spectral index of the pulsar radio emission, $\nu_0\approx 10^8$ Hz,
\begin{equation}
I_{\nu_0}=\frac{L}{S\nu_0},
\label{eq38}
\end{equation}
$L$ is the total radio luminosity of the pulsar, $S=\pi r^2\vartheta^2/4$ is the cross-section of the radio emission cone at a distance $r$ and $\vartheta$ is the pulse width in the angular measure. Then
\[
\Gamma=900\frac{L}{10^{29}\,\mathrm{erg\,s}^{-1}}\frac{0.1\,\mathrm{s}}{P}\frac{B_\star}{10^{12}\,\mathrm{G}}\left (\frac{\vartheta}{0.1}\right )^{-2}\left (\frac{r}{10^8\,\mathrm{cm}}\right )^{-4}
\]
\begin{equation}
\times\left (\frac{\nu^\prime}{10^9\,\mathrm{Hz}}\right )^{-2}\left (\frac{\nu}{10^8\,\mathrm{Hz}}\right )^{-\alpha}\frac{\kappa}{10^2}\frac{10^3}{\gamma}\left (\frac{\gamma_0}{10}\right )^{-2}.
\label{eq39}
\end{equation}
Taking $L=2\times 10^{29}$ erg\,s$^{-1}$, $\vartheta=0.15$, $r=3\times 10^8$ cm as well as $P$ and $B_\star$ listed above, we obtain $\Gamma=30$. Thus, in the Vela pulsar the induced scattering of the radio beam into the background can indeed be efficient. Furthermore, in the course of pulse-to-pulse fluctuations of the parameters, the induced scattering can from time to time enter the stage of saturation, when the original radio beam is substantially suppressed (see Fig.~\ref{f3}).

It should be kept in mind that the pulsar radio beam is broadband and its angle of incidence increases with distance from the neutron star, $\theta\propto r$ (see below). Therefore at different altitudes $r$ the background component of a given frequency $\nu^\prime=\nu\theta^2(r)\gamma^2/2\gamma_0^2$ is fed by the beam radiation of different frequencies $\nu\propto r^{-2}$. As at lower frequencies the beam radiation is more intense, the induced scattering is more efficient at higher altitudes above the neutron star \citep[for more detail see][]{p08b}. Thus, the effective scattering region lies far from the emission region, at distances of the order of the cyclotron resonance and the light cylinder radii, $r\sim r_c,r_L$.

\begin{figure*}
\centering
\includegraphics{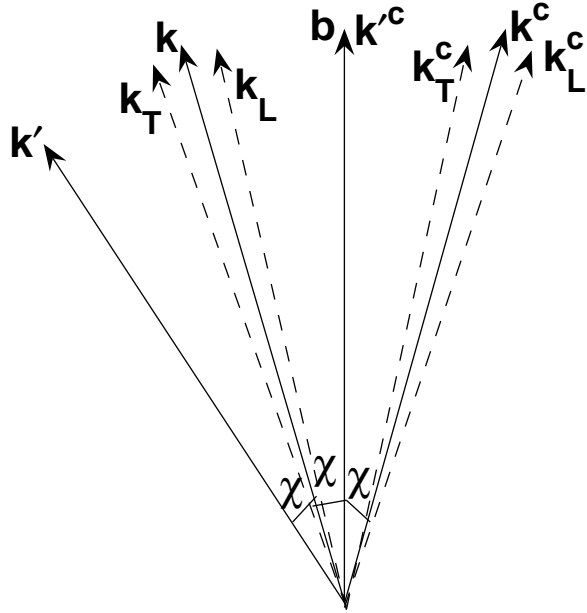}
\caption{Geometry of induced scattering at the altitude $r$ in the rotating magnetosphere. The rotation is counterclockwise, the superscript 'c' marks the wavevectors in the frame corotating with the neutron star, the dashed arrows show the orientations of the wavevectors at the leading (L) and trailing (T) edges of the pulse with respect to the magnetic field direction at the point of scattering; $\chi\equiv r/2r_L$.}
\label{f4}
\protect\label{lastpage}
\end{figure*}

The location of the scattered component in the pulse profile is predominantly determined by the effect of the magnetosphere rotation \citep{p08b}. In the scattering region, the incident beam makes the angle $\sim r/2r_L$ with the local magnetic field direction $\vec{b}$. Note that in the frame corotating with the neutron star the beam is shifted in by $r/r_L$ oppositely to the direction of rotation and again makes the angle $r/2r_L$ with the magnetic field (see Fig.~\ref{f4}). As for the scattered component, in the corotating frame it is directed approximately along the magnetic field. Hence, in the pulse profile it precedes the original radio beam by $\Delta\lambda=r/2r_L$ in phase. Thus, the induced scattering of the pulsar beam into the background gives rise to the leading component of the profile, which is usually called the precursor.

To analyze the radio profile structure of the Vela pulsar in more detail it is necessary to take into account the finitude of the pulse width. In the scattering region, different radio beams (which constitute the emission cone and appear in the pulse profile at different pulse phases) make somewhat different angles with the local magnetic field: the angle of incidence $\theta$ ranges roughly from $r/2r_L-\vartheta/4$ to $r/2r_L+\vartheta/4$. 
For a fixed frequency of the incident radiation, $\nu$, the scattering efficiency $\Gamma$ depends on $\theta$: $\Gamma\propto\nu^{\prime -2}\propto\theta^{-4}$, and therefore different parts of the main pulse suffer the scattering of various strength. The angle $\theta$ entering Eq.~(\ref{eq39}) corresponds to the corotating frame, and, as can be seen from Fig.~\ref{f4}, smaller values of $\theta$ are characteristic of the rays forming the trailing part of the profile. Thus, it is the trailing part that should be scattered more efficiently.

For the radio pulse passing through the scattering region the assumption of the stationary scattering seems reliable. At the same time, the incident intensity and the parameters of the scattering plasma are expected to fluctuate from pulse to pulse. The resultant fluctuations of $\Gamma$ affect not only the observed radio profile shape but also the total intensity of the profile. Recall that the induced scattering leads to the intensity transfer between widely spaced frequencies, $\nu^\prime=\nu\theta^2\gamma_\Vert^2/2\gg\nu$, and a substantial part of the lower-frequency intensity $I_\nu^{(0)}$ may come to the higher frequency (cf. Eq.~(\ref{eq32})). With the decreasing spectrum of the pulsar radio emission, this may imply a noticeable increase of the total intensity of the higher-frequency profile due to the strong precursor component resulting from the scattering. Thus, the induced scattering increases the radio profile intensity at a fixed frequency, the amplification being stronger for higher $\Gamma$.

All this is in line with the observed properties of the Vela's radio pulses \citep{kd83}. A more pronounced precursor component is indeed characteristic of stronger pulses. At the same time, the rest of the profile weakens with the total intensity growth, being more efficiently suppressed by the induced scattering to higher frequencies. In stronger pulses, where the role of induced scattering is more significant, the region of substantial suppression extends to earlier pulse phases, where $\Gamma$ becomes large enough.

It should be noted that for any angle of incidence of the radio beam the orientation of the scattered component is the same, $\theta^\prime\approx 1/\gamma_\Vert$, and therefore the shape of the precursor component should be determined by the variation of the magnetic field orientation and the scattering efficiency across the scattering region. The magnetic field direction changes monotonically, so that the leading edge of the precursor is formed by the radiation at the leading edge of the main pulse and vice versa. For a fixed frequency of the scattered radiation $\nu^\prime$ we have $\nu\theta^2=$const, i.e. at different longitudes the precursor component is fed by the main pulse radiation of different frequencies, and the efficiency of the component growth $\Gamma\propto\nu^{-\alpha}\propto\theta^{2\alpha}$. As the angle of incidence is larger for the leading rays (see Fig.~\ref{f4}), the peak of the precursor is formed by the radiation at the extreme leading edge of the main pulse, where the original intensity is well below the profile maximum but is large enough to provide an efficient scattering. Hence, in stronger pulses, where the scattering efficiency is higher, a significant contribution to the scattered component comes from the main pulse rays at the earlier pulse longitudes and, correspondingly, the precursor arrives somewhat earlier.
The observations do reveal the unambiguous relation between the arrival time of the pulse and its total intensity \citep{kd83}. In our model, both quantities are determined by the efficiency of induced scattering, and the salient features of the radio profile structure and its fluctuations are explained naturally.

The frequency evolution of the pulse profile is also worthy to be addressed. According to Eq.~(\ref{eq39}), the scattering efficiency is much larger at lower frequencies. Therefore the precursor component should be most pronounced at lower frequencies, whereas the rest of the pulse should dominate at higher frequencies. This is proved by the observational data. There is only a hint of a component on the trailing edge of the pulse at 400 MHz \citep{hamilton77}, at 2.295 GHz it becomes more convincing \citep{downs79,kd83}, and at 4.8 GHz the trailing component is stronger than the precursor \citep{Gardner69}.

The radio observations of \citet{l07}, which have revealed the correlation with the high-energy emission, have been carried out at the frequency of 1 GHz. Unfortunately, in the above mentioned paper the radio profile properties have not been presented in detail and the authors have concentrated solely on the time of arrival of the radio pulses, assuming implicitly that the basic features of the radio profile structure and its variations with the radio pulse intensity are the same as those described in \citet{kd83} for the data at 2.3 GHz.

\subsection{High-energy emission as a result of radio photon reprocessing \label{s4.2}}

Let us turn to the consequences of both the spontaneous scattering of the radio pulse and the particle synchrotron re-emission in application to the Vela pulsar. As is pointed out above, our consideration is restricted to the scattering of the A-polarization, since only this mode is present in the Vela's radio emission. The characteristic frequency of the scattered radiation is given by $\omega_{\mathrm{max}}^\prime\eta^\prime=\omega\eta+s_{\mathrm{max}}\Omega\approx s_{\mathrm{max}}\Omega$, where $\eta^\prime=1/\gamma_\Vert^2$, $s_{\mathrm{max}}=3y_{\mathrm{max}}\gamma_0^3/2$ is the harmonic number corresponding to the spectral maximum of the scattered radiation and $y_{\mathrm{max}}=5/4$. This is reduced to $\omega^\prime=1.5y_{\mathrm{max}}\Omega\gamma^2\gamma_0$ and can be estimated as
\begin{equation}
\hbar\omega^\prime=1.7\times 10^2y_{\mathrm{max}}\frac{B_\star}{10^{12}\,\mathrm{G}}\frac{\gamma}{10^3}\frac{\gamma_0}{10}\left (\frac{r}{10^8\,\mathrm{cm}}\right )^{-3}\,\mathrm{eV}.
\label{eq40}
\end{equation}
The characteristic frequency of the re-emitted radiation is given by Eq.~(\ref{eq40}) with $y_{\mathrm{max}}=0.3$. Given that $B_\star=3.4\times 10^{12}$ G we find $\hbar\omega_{\mathrm{max}}=0.8$ keV and 0.2 keV for the scattered and re-emitted radiation, respectively. Thus, the spectrum of the reprocessed radiation can extend up to the soft X-ray range.

Proceeding from Eq.~(\ref{eq23}), which gives the scattered and synchrotron powers of a single particle, one can estimate the luminosities due to the scattering and re-emission, $L_{\mathrm{sc}}=(P^{AA}+P^{AB})N_eSr$ and $L_{\mathrm{syn}}=(P_{\mathrm{syn}}^A+P_{\mathrm{syn}}^B)N_eSr$. Using Eqs.~(\ref{eq35}), (\ref{eq37}) and (\ref{eq38}), we find
\[
L_{\mathrm{sc}}=2\times 10^{26}\left (\frac{B_\star}{10^{12}\,\mathrm{G}}\frac{0.1\,\mathrm{s}}{P}\right )^5\frac{L}{10^{29}\,\mathrm{erg\,s}^{-1}}\left (\frac{\nu}{10^9\,\mathrm{Hz}}\right )^{-4-\alpha}
\]
\begin{equation}
\times\frac{\kappa}{10^2}\left (\frac{\gamma}{10^3}\right )^{-4}\left (\frac{\gamma_0}{10}\right )^6\left (\frac{r}{10^8\,\mathrm{cm}}\right )^{-20}\, \mathrm{erg\,s^{-1}},
\label{eq41}
\end{equation}
\[
L_{\mathrm{syn}}=10^{29}\left (\frac{B_\star}{10^{12}\,\mathrm{G}}\right )\left (\frac{P}{0.1\,\mathrm{s}}\right )^{-1}\left (\frac{\vartheta}{0.1}\right )^2
\]
\begin{equation}
\times\frac{\kappa}{10^2}\left (\frac{\gamma_0}{10}\right )^2\left (\frac{r}{10^8\,\mathrm{cm}}\right )\,\mathrm{erg\,s^{-1}}.
\label{eq42}
\end{equation}
Recall that $L$ is the total radio luminosity of the pulsar, $\alpha$ is the spectral index of the radio emission ($\alpha=1.3$ for $\nu<10^9$ Hz and $\alpha=2.7$ for $\nu>10^9$ Hz). Taking into account the parameters of the Vela pulsar, $L=2\times 10^{29}$ erg\,s$^{-1}$, $P=0.089$ s, $B_\star=3.4\times 10^{12}$ G and $\vartheta=0.15$, yields $L_{\mathrm{sc}}\sim 10^{29}$ erg\,s$^{-1}$ and $L_{\mathrm{syn}}\sim 10^{31}$ erg\,s$^{-1}$. The latter value is consistent with the observed luminosity of the Vela pulsar in the range of 0.2--8 keV. \citep{pavlov_01}.

Our technique developed in Sect.~2 implies that the scattering is considered as a secondary process, and it is applicable as long as the total power scattered is less than the synchrotron power, i.e. $L_\mathrm{sc}/L_\mathrm{syn}\ll 1$. Comparison of Eqs.~(\ref{eq41}) and (\ref{eq42}) shows that this inequality is generally valid, though both luminosities are strong functions of the parameters. The role of the scattered emission is believed to be most significant in the range of a few keV, where the spectral power of the scattered radiation reaches maximum, while the synchrotron spectrum drops exponentially (see Fig.~\ref{f1}). 

It is important to note that the luminosity due to the scattering depends on the Lorentz-factor of the particle gyration much stronger than the synchrotron luminosity: $L_\mathrm{sc}\propto\gamma_0^6$, whereas $L_\mathrm{syn}\propto\gamma_0^2$. The evolution of the particle Lorentz-factor $\gamma_0$ is determined by the radio intensity, which comes to the resonance region and causes the particle momentum evolution. Then it is the scattered luminosity $L_\mathrm{sc}$ that is strongly sensitive to the radio emission characteristics and their fluctuations. Hence, the high-energy emission is believed to exhibit most pronounced correlation with radio emission in the range beyond the synchrotron maximum, where the contribution of the scattered power is substantial. This is indeed observed in the Vela pulsar.

It should be noted that the particle distribution function in momenta is sufficiently broad, and it is believed to evolve with distance significantly. Therefore the resultant synchrotron spectrum do not resemble the single-particle spectrum: due to the strong spatial dependence of $\omega_\mathrm{max}^\prime$ the spectral maximum is smeared, and the total spectrum of the particles is modified substantially \citep[for a more detailed analysis see][]{p03}. The same is expected for the scattered spectrum as well. Therefore in the present paper we do not analyze the spectral slopes.

Now let us consider the location of the scattered and re-emitted radiation in the pulse profile. Both components of the high-energy emission resulting from the radio photon reprocessing are believed to arise at an angle $\sim 1/\gamma_\Vert$ to the ambient magnetic field. Then the high-energy emission generated close enough to the radio emission region should appear in the profile at the original radio pulse position. The high-energy emission originating at higher altitudes should appear $\sim r/2r_L$ ahead of the main pulse in radio (see Sect.~4.1). In particular, the radiation generated at the altitudes of the order of the cyclotron resonance radius, $r\sim r_c$, should coincide in phase with the radio precursor position. The emission region of the high-energy component which results from the scattering of the under-resonance radio photons is restricted to $r\sim r_c$, and, correspondingly, this component is present only in the same phase range as the radio precursor. This range is usually classified as the position of peak 4 in the high-energy profile \citep{h02}. As for the synchrotron emission, it can be efficient over a wider range of altitudes, even beyond the resonance region, and it is expected to contribute significantly to both peak 4 and peak 3, which precedes the radio pulse by $\sim 30^\circ$. Note that the position of peak 3 implies the component origin close to the light cylinder. It is important to point out that, in contrast to the rest of the high-energy profile, peaks 3 and 4 are present only in the optical and soft X-ray range, up to a few keV, where they turn into the trough. All this strongly supports an idea of their origin as a result of synchrotron re-emission and spontaneous scattering by spiraling particles.

Peak 1 of the high-energy profile is also of interest, since its intensity is also affected by the radio pulse properties \citep{l07}. It should be noted that in the optical -- soft X-ray range peak 1 noticeably shifts with frequency toward the radio pulse location, whereas at higher energies it keeps a fixed position \citep{h02}. The nature of peak 1 seems questionable. Its position in the pulse profile, $\sim 90^\circ$ after the radio pulse, excludes the magnetospheric origin of this component provided that the high-energy emission is directed approximately along the magnetic field. One can speculate, however, that the radiation forming peak 1 presents the synchrotron emission of the particles beyond the light cylinder. A more detailed analysis of the high-energy profile of the Vela pulsar is beyond the framework of the present paper.

\subsection{Manifestations of the radio -- high-energy connection \label{s4.3}}

The physical connection of the high-energy radiation considered in Sect.~4.2 to the radio pulse pulse properties discussed in Sect.~4.1 is expected to have observational manifestations. To analyse them let us first note that the spontaneous scattering into the high-energy range is most efficient at low enough altitudes, $r\ll r_c$ (cf. Eq.~(\ref{eq41})), due to larger number densities of the scattering particles, $N_e\propto r^{-3}$, and larger incident intensities, $I_\nu\propto r^{-2}$, in this region. The radio precursor component is formed at higher altitudes, $r\sim r_c$, where the rate of spontaneous scattering is already less. Besides that, the precursor arises at a small angle $\theta^\prime\sim 1/\gamma_\Vert$ to the ambient magnetic field, and, correspondingly, the scattering cross-section for the incident A-polarization is less (cf. Eq.~(\ref{eq23})). In addition, the orientation of the precursor with respect to the ambient magnetic field rapidly changes to $\theta^\prime=r/2r_L\gg 1/\gamma_\Vert$, so that the original under-resonance radiation, $\omega^\prime/\gamma_\Vert^2\ll\Omega$, passes through the resonance, $\omega^\prime\theta^{\prime^2}/2=\Omega$, very soon. Therefore it is the main pulse that chiefly contributes to the spontaneous scattering to high energies. In contrast to the resonant absorption, this process does not affect the radio pulse intensity considerably \citep{p08a}. Note also that the resonant absorption suppresses the main pulse and precursor alike, so that the synchrotron re-emission is related to both radio components equally.

Thus, the main pulse of the radio profile is subject to both the spontaneous and induced scatterings. The first process contributes to the soft X-ray component roughly coincident with the radio pulse window, whereas the second process gives rise to the precursor component on the radio profile. The efficiencies of the two processes are expected to vary from pulse to pulse because of fluctuations of the plasma parameters. This is believed to underlie the observational manifestation of the X-ray -- radio connection in the Vela pulsar. Weaker induced scattering implies less efficient intensity transfer from the main pulse to the precursor. Then the resultant intensity of the radio pulse is less (see Sect.~4.1), the precursor is weaker and the main pulse is less suppressed. In this case, the main pulse can more efficiently participate in the spontaneous scattering, and the resultant high-energy component should be more pronounced. This is in line with the observed trend: in the Vela pulsar weaker radio pulses are accompanied by stronger high-energy emission at the position of peaks 3 and 4 \citep{l07}. One can speculate that the fluctuations of the effective value of $\gamma_0$ make the dominant contribution to the variations of the efficiencies of the spontaneous and induced scatterings. At larger $\gamma_0$ the induced scattering is weaker (see Eq.~(\ref{eq39})) and the high-energy luminosities $L_\mathrm{sc}$ and $L_\mathrm{syn}$ are larger (see Eqs.~(\ref{eq41})-(\ref{eq42})).

Less values of $\gamma_0$ mean weaker momentum evolution of the particles in the course of resonant absorption of radio emission. Then a greater part of the particle gyration energy is expected to be re-emitted at higher altitudes, beyond the light cylinder, contributing to peak 1 of the high-energy profile. Hence, strong radio pulses with more pronounced precursors, which correspond to large $\Gamma$, should be accompanied by stronger high-energy emission at the position of peak 1 and weaker emission in the trough. This also agrees with the observed trend.

\section{Summary and discussion \protect\label{s5}}

We have considered the processes of spontaneous and induced scattering off the particles performing relativistic helical motion in an external magnetic field. The theory is applied to the radio wave scattering off the secondary plasma particles in the pulsar magnetosphere. The particles are believed to have substantial gyration energies due to resonant absorption of the radio emission in the outer magnetosphere.

In application to the Vela pulsar, the induced scattering of radio waves between the states well below the resonance, $\omega\eta=\omega^\prime\eta^\prime\ll\Omega$, can be efficient. An extremely bright and narrow radio beam is scattered chiefly into the background, in the direction corresponding to the maximum scattering probability. The scattered component is directed approximately along the ambient magnetic field, $\theta_\mathrm{max}^\prime\sim 1/\gamma_\Vert$, and appears in the radio profile as a precursor to the main pulse. This scenario for the first time allows to explain the main features of the radio profile structure of the Vela pulsar and its pulse-to-pulse fluctuations.

In the case considered, the induced scattering transfers the main pulse intensity to the higher frequencies, $\omega^\prime\sim\omega\theta^2\gamma_\Vert^2\gg\omega$. With the decreasing spectrum of the pulsar radio emission, this implies intensity enhancement of the radio profile at a fixed frequency. Thus, larger scattering efficiencies should result in both higher pulse intensities and more pronounced precursors. Such a correlation is really observed in the Vela pulsar \citep{kd83}. Moreover, stronger scattering means that a larger part of the main pulse can be scattered efficiently, the region of substantial suppression extending from the very trailing edge of the profile toward earlier phases. As the scattering of the leading edge of the main pulse makes the dominant contribution to the precursor formation, higher scattering efficiencies imply not only larger intensities but also earlier phases of the precursor peak. The observations do reveal early arrival of the precursor component in strong pulses \citep{kd83}.

The radio pulse is also subject to spontaneous scattering by the spiraling particles. The photons below the resonance, $\omega\eta\ll\Omega$, are chiefly scattered to high harmonics of the particle gyrofrequency, $s\sim\gamma_0^3$. We have analyzed the characteristics of the scattered radiation in detail and compared them with those of the synchrotron radiation of the same particle. In particular, it is found that the scattered power peaks at somewhat higher energies than the synchrotron one. In application to the Vela pulsar, the estimates of the spectral maxima yield $\sim 0.8$ keV and $\sim 0.2$ keV, respectively. The synchrotron luminosity is $\sim 10^{31}$ erg\,s$^{-1}$ and well agrees with the observed values in the soft X-ray band \citep[e.g][]{pavlov_01}. Although the total luminosity provided by the scattering is less, in the range beyond the synchrotron maximum the scattered power may still contribute substantially. It is important to note an extremely strong dependence of the scattered power on the Lorentz-factor of the particle gyration, $L_\mathrm{sc}\propto\gamma_0^6$, which means a strong relation of the scattered component to the radio intensity.

Similarly to the synchrotron emission, the scattered radiation concentrates close to the ambient magnetic field direction, $\theta^\prime\sim 1/\gamma_\Vert$. Given that the synchrotron and scattered emissions originate well above the radio emission region and inside the light cylinder, they should precede the main radio pulse by $\la 30^\circ$. This position can be identified with that of peaks 3 and 4 of the soft X-ray profile, which turn into the trough at somewhat higher energies \citep[e.g.][]{h02}. If the synchrotron re-emission continues beyond the light cylinder, it may contribute to peak 1 of the profile.

The radio emission of the Vela pulsar is believed to participate in both the spontaneous and induced scatterings, and the interplay between these processes can account for the observed X-ray -- radio correlation. If the main pulse is less suppressed by the induced scattering to the precursor, it is more efficiently scattered to high energies. This is consistent with the observations: weaker radio pulses with less pronounced precursors are accompanied by the high-energy pulses with stronger emission in the trough \citep{l07}. Given that the variations of the scattering efficiencies are determined by the fluctuations of $\gamma_0$, larger $\gamma_0$ imply less efficient induced scattering and simultaneously stronger high-energy luminosities. For smaller $\gamma_0$ the synchrotron re-emission is expected to continue beyond the light cylinder, contributing to peak 1 of the high-energy profile. At the same time, smaller $\gamma_0$ imply more efficient induced scattering, a more pronounced radio precursor and stronger resultant radio pulses. All this is in line with the observed trends \citep{kd83,l07}. Thus, our model explains the salient features of the radio profile formation of the Vela pulsar, the peculiarities of its soft X-ray profile as well as the observed X-ray -- radio connection.

It should be noted that our model of the radio profile formation in the Vela pulsar is too simplified, since it includes only two components, the main pulse and precursor. \citet{kd83} have established four components of the Vela's radio profile. Later on \citet{j01} and \citet{j02} have discovered sporadic activity at the leading edge of the pulse (the so-called giant micro pulses) and in the bump region in the trailing part of the profile. Further development of our model is needed in order to include these phenomena. However, we believe that these peculiarities do not affect the X-ray -- radio connection noticeably.

We have considered the induced scattering in the approximation of a strong magnetic field. This process can be efficient only if the incident and scattered waves have the ordinary polarization (the A-polarization), i.e. if their electric vectors are in the plane of the ambient magnetic field. In the Vela pulsar, only one polarization mode is present. Various emission theories used to identify it with the ordinary mode. The point is that the extraordinary (B-) mode has the vacuum dispersion, and its direct generation by any plasma mechanism seems problematic. The recent high-energy observations of the Vela's pulsar wind nebula have cast some doubt as to the type of the radio polarization in this pulsar \citep{jet1,jet2,jet3}. They have revealed the jet in the direction of the pulsar proper motion. If this jet is directed along the pulsar rotational axis, then the radio emission has extraordinary polarization. Note that jet alignment with the rotational axis is well ascertained for the accretion systems, but the radio pulsars do not seem to be such systems, and the nature of the Vela's jet is obscure. Therefore we have still assumed the ordinary polarization of the Vela's radio emission.

The high-energy emission of the Vela pulsar is too complicated, and the emission mechanisms involved cannot be exhausted by those considered in the present paper. As is argued by \citet{h02}, the Vela's high-energy emission consists of the soft and hard components. The radio photon reprocessing contributes solely to the soft component. The photons emitted inside the light cylinder appear in the phase region preceding the radio pulse by $\la 30^\circ$. The synchrotron re-emission beyond the light cylinder may at least partially contribute to peak 1, whereas peak 2 is thought to result from some other mechanism.

Within the framework of our model, the peculiar X-ray -- radio correlation observed in the Vela pulsar is only characteristic of a narrow spectral range in the soft X-ray band, namely just beyond the spectral maximum of the high-energy component roughly coincident with the radio pulse window. It is the range where the contribution of the scattering off the spiraling particles may be substantial, and the scattered component is tightly connected to the radio pulse intensity. At the same time, another process of the radio photon reprocessing to high energies -- the synchrotron re-emission of the spiraling particles -- is believed to be significant over a wider spectral range. Further observational studies of the X-ray -- radio correlation at softer energies would be of interest. A more detailed comparison of the high-energy data with the radio profile properties would be particularly useful. 

The radio photon reprocessing is believed to take place in other pulsars as well. The resultant radio -- high-energy connection is yet to be discovered. It should be kept in mind that the observational manifestations of this connection may be quite different because of the difference of the physical conditions and parameters. In the radio range, pulsars are known to exhibit various fluctuation phenomena, and the simultaneous observations in the radio and soft high-energy ranges are expected to reveal diversiform connections.

{}

\begin{appendix}
\section{Integration of the components of the scattered power $P^{AA}$ and $P^{AB}$ over the angle \label{sa}}

We are going to calculate the integrals
\begin{equation}
S_1\equiv\int\limits_{-\infty}^\infty\psi^4(1+\psi^2)K_{1/3}^2\left [\frac{y}{2}(1+\psi^2)^{3/2}\right ]\mathrm{d}\psi
\label{a1}
\end{equation}
and
\begin{equation}
S_2\equiv\int\limits_{-\infty}^\infty\psi^2(1+\psi^2)^2K_{2/3}^2\left [\frac{y}{2}(1+\psi^2)^{3/2}\right ]\mathrm{d}\psi.
\label{a2}
\end{equation}
Following the technique developed in \citet{w59}, we proceed from the integral representation of $K_{1/3}(\zeta)$,
\begin{equation}
K_{1/3}(\zeta)=\frac{3}{2\sqrt{\mu q}}\int\limits_{-\infty}^\infty\mathrm{e}^{\mathrm{i}\left (t^3+\mu qt\right )}\mathrm{d}t,
\label{a3}
\end{equation}
where $\zeta=2(\mu q/3)^{3/2}$, $\mu=3y^{2/3}/2^{4/3}$ and $q=1+\psi^2$. Then one can write
\begin{equation}
qK_{1/3}^2(\zeta)=\frac{9}{4\mu}\int\limits_{-\infty}^\infty\int\limits_{-\infty}^\infty
\mathrm{e}^{\mathrm{i}(q\mu u+u^3)}{e}^{-\mathrm{i}(q\mu v+v^3)}\mathrm{d}u\mathrm{d}v.
\label{a4}
\end{equation}
In terms of the new variables,
\begin{equation}
x=2^{-2/3}(u-v)\quad \mathrm{and}\quad\xi=2^{-2/3}(u+v),
\label{a5}
\end{equation}
Eq.~(\ref{a4}) can be integrated over $\xi$ taking into account that
\begin{equation}
\int\limits_{-\infty}^\infty\xi^{2n}\mathrm{e}^{\mathrm{i}\alpha\xi^2}\mathrm{d}\xi=\frac{(2n-1)!!}{2^n}\sqrt{\pi}\mathrm{e}^{\mathrm{i}\pi/4}\mathrm{e}^{\mathrm{i}n\pi/2}\alpha^{-n-1/2}.
\label{a6}
\end{equation}
Then
\[
S_1=\frac{9\sqrt{\pi}\mathrm{e}^{\mathrm{i}\pi/4}}{2^{5/3}\mu\sqrt{3}}\int\limits_{-\infty}^\infty\int\limits_{-\infty}^\infty\psi^4\mathrm{e}^{\mathrm{i} 2^{2/3}(\psi^2+1)\mu x+\mathrm{i} x^3}\mathrm{d}\psi x^{-1/2}\mathrm{d}x.
\]
Integration over $\psi$ with the help of Eq.~(\ref{a6}) yields
\begin{equation}
S_1=\frac{-27\pi\mathrm{i}}{2^{16/3}\sqrt{3}\mu^{7/2}}\int\limits_{-\infty}^\infty x^{-3}\mathrm{e}^{\mathrm{i} 2^{2/3}\mu x+\mathrm{i} x^3}\mathrm{d}x.
\label{a7}
\end{equation}
This can be integrated by parts,
\begin{equation}
\int\limits_{-\infty}^\infty x^{-3}\mathrm{e}^{\mathrm{i}2^{2/3}\mu x+\mathrm{i}x^3}\mathrm{d}x=\int\limits_{-\infty}^\infty\frac{\mathrm{i}2^{2/3}\mu+3\mathrm{i}x^2}{2x^2}\mathrm{e}^{\mathrm{i}2^{2/3}\mu x+\mathrm{i}x^3}\mathrm{d}x.
\label{a8}
\end{equation}
Then the second term on the right-hand side is directly related to $K_{1/3}(\zeta)$ (see Eq.~(\ref{a3})), and the first one can be integrated by parts once more,
\begin{equation}
\int\limits_{-\infty}^\infty x^{-2}\mathrm{e}^{\mathrm{i}2^{2/3}\mu x+\mathrm{i}x^3}\mathrm{d}x=\int\limits_{-\infty}^\infty\frac{\mathrm{i}\mu 2^{2/3}+3\mathrm{i}x^2}{x}\mathrm{e}^{\mathrm{i}2^{2/3}\mu x+\mathrm{i}x^3}\mathrm{d}x.
\label{a9}
\end{equation}
The second term in Eq.~(\ref{a9}) is related to $\partial/\partial\mu\{\sqrt{\mu}K_{1/3}[4(\mu/3)^{3/2}]\}$ (cf. Eq.~(\ref{a3})), whereas the first one can be obtained by integrating $\int_\mu^\infty\sqrt{\mu}K_{1/3}[4(\mu/3)^{3/2}]\mathrm{d}\mu$. Then, making use of the recurrence relation
\begin{equation}
-K_\nu^\prime(\xi)=K_{\nu-1}(\xi)+\frac{\nu}{\xi}K_\nu(\xi)
\label{a10}
\end{equation}
and keeping in mind that $K_\nu(\xi)=K_{-\nu}(\xi)$, one can obtain
\begin{equation}
S_1=\frac{3\pi}{4\sqrt{3}y}\left [\frac{2}{3y}K_{1/3}(y)-K_{2/3}(y)+\int\limits_y^\infty K_{1/3}(x)\mathrm{d}x\right ].
\label{a11}
\end{equation}
With the recurrence relations
\begin{equation}
-2K_\nu^\prime(\xi)=K_{\nu-1}(\xi)+K_{\nu+1}(\xi),
\label{a12}
\end{equation}
\begin{equation}
-\frac{2\nu}{\xi}K_\nu(\xi)=K_{\nu-1}(\xi)-K_{\nu+1}(\xi),
\label{a13}
\end{equation}
this is reduced to
\begin{equation}
S_1=\frac{3\pi}{4\sqrt{3}y}\left [K_{4/3}(y)-\int\limits_y^\infty K_{5/3}(x)\mathrm{d}x\right ].
\label{a14}
\end{equation}

The integral $S_2$ can be treated analogously. From Eqs.~(\ref{a10}) and (\ref{a3}) one can find that
\begin{equation}
\frac{q\mu^{3/2}}{\sqrt{3}}K_{2/3}(\zeta)=-\frac{3}{2\sqrt{\mu}}\frac{\partial}{\partial q}\int\limits_{-\infty}^\infty\mathrm{e}^{\mathrm{i}q\mu t+\mathrm{i}t^3}\mathrm{d}t.
\label{a15}
\end{equation}
Then
\begin{equation}
q^2K_{2/3}^2(\zeta)=\frac{27}{4\mu^2}\int\limits_{-\infty}^\infty\int\limits_{-\infty}^\infty\mathrm{e}^{\mathrm{i}q\mu u+\mathrm{i}u^3}\mathrm{e}^{-\mathrm{i}q\mu v-\mathrm{i}v^3}uv\mathrm{d}u\mathrm{d}v.
\label{a16} 
\end{equation}
With the variables given by Eq.~(\ref{a15}), this is reduced to
\[
q^2K_{2/3}^2(\zeta)=\frac{27\pi}{2^{7/3}\mu^2}\int\limits_{-\infty}^\infty\mathrm{e}^{\mathrm{i}2^{2/3}q\mu x+\mathrm{i}x^3}\mathrm{d}x\int\limits_{-\infty}^\infty(\xi^2-x^2)\mathrm{e}^{\mathrm{i}3x\xi^2}\mathrm{d}\xi
\]
and can be integrated over $\xi$ with the help of Eq.~(\ref{a6}). Then $S_2$ can be written as
\begin{equation}
S_2=\frac{-27\pi}{2^{13/3}\sqrt{3}\mu^{7/2}}\int\limits_{-\infty}^\infty\left (\frac{\mathrm{i}}{6x^3}-1\right )\mathrm{e}^{\mathrm{i}2^{2/3}\mu x+\mathrm{i}x^3}\mathrm{d}x,
\label{a17}
\end{equation}
where the integration over $\psi$ is also performed using Eq.~(\ref{a6}). One can see that the first term of $S_2$ is proportional to that given by Eq.~(\ref{a8}), whereas the second one can be expressed in terms of the function $K_{1/3}(\zeta)$ (cf. Eq.~(\ref{a3})). Thus, one can find finally
\begin{equation}
S_2=\frac{\pi}{4\sqrt{3}y}\left [\frac{8}{3y}K_{1/3}(y)+K_{4/3}(y)-\int\limits_y^\infty K_{5/3}(x)\mathrm{d}x\right ].
\label{a18}
\end{equation}

\end{appendix}

\clearpage

\clearpage

\clearpage

\clearpage

\protect\label{lastpage}

\end{document}